# MESSAGING WITH COST OPTIMIZED INTERSTELLAR BEACONS

James Benford
Microwave Sciences, Inc. USA
Lafayette CA 94549 USA
Phone: 925-283-8454 Fax: 925-283-8487
jbenford@earthlink.net

Gregory Benford
Physics and Astronomy Dept., UC Irvine USA
University of California Irvine, Irvine CA 92612 USA

Dominic Benford
Observational Cosmology Lab, NASA/Goddard Space Flight Center, Greenbelt MD 20771 USA

How would we on Earth would build galactic-scale Beacons to attract the attention of extraterrestrials, as some have suggested we should do? From the point of view of expense to a builder on Earth, experience shows an optimum tradeoff. This emerges by minimizing the cost of producing a desired power density at long range, which determines the maximum range of detectability of a transmitted signal. We derive general relations for cost-optimal aperture and power. For linear dependence of capital cost on transmitter power and antenna area, minimum capital cost occurs when the cost is equally divided between antenna gain and radiated power. For non-linear power law dependence a similar simple division occurs. This is validated in cost data for many systems; industry uses this cost optimum as a rule-of-thumb. Costs of pulsed cost-efficient transmitters are estimated from these relations using current cost parameters ($/W, $/m$^2$) as a basis. We show the scaling and give examples of such Beacons. Galactic-scale Beacons can be built for a few billion dollars with our present technology. Such beacons have narrow 'searchlight' beams and short 'dwell times' when the Beacon would be seen by an alien observer in their sky. On this cost basis they will likely transmit at higher microwave frequencies, ~10 GHz. The natural corridor to broadcast is along the galactic radius or along the local spiral galactic arm we are in. A companion paper, asks 'If someone like us were to produce a Beacon, how should we look for it?'

1. **Introduction**

Sending signals into space in the hope that they will be received by an alien intelligence is called Messaging to Extraterrestrial Intelligence (METI) or Active SETI. This is the first of two papers to consider the proposal quantatively. Here we describe how we would make beacons for someone else to see. In our second paper, "Cost Optimized Interstellar Beacons: METI", we show, if someone like us were to produce a Beacon, here's what we should look for.

The handful of small-scale transmissions done to date use radio telescopes and radiate <MW in the microwave to nearby stars. Reaching out on large scales > 1000 ly. demands far more power and expense. Economics will matter crucially.

Beyond ~1000 light years, interstellar obscuration makes identifying telltale biological features such as an ozone spectral line difficult. This means a METI program would sweep regions of the sky, not target individual stars. METI ranges > 1000 light years require an Effective Isotropic Radiated Power (EIRP) >$10^{17}$ W, so that the broadcaster enters the domain not of targeted radiators, but of Beacons. The most advanced case is that of microwave Beacons for communication across galactic distances, using EIRPs of $10^{17}$-$10^{20}$ W. For comparison, the Arecibo radio telescope EIRP is ~$10^{13}$ W. METI messages have been sent at EIRPs of $10^{12}$-$10^{13}$ W (Zatsev, 2006), so have short ranges.

Advocates of METI argue there are several motivations given for building Beacons:
1) As Earth becomes more electromagnetically quiet, driven by fiber-optic communication, we should make ourselves more visible, to aid in others in races finding us;
2) We can better understand what we're looking for in SETI if we develop our own concepts for transmitting large powers into space.

There is controversy about whether we should transmit and about who should decide (Brin, 2007). The objections to METI are that it poses a danger, that contact with a benign extraterrestrial civilizations would mean serious societal stress for Earth, and that not all such civilizations will be benign.

Here we address quantitative questions: How could a long-range Earth-based METI transmitter be built? What power would it have? How big would it be? What would it cost? What radiating strategy should it follow? We find that these questions are interlinked, strongly related to each other. They also give insights into what SETI searches should be looking for.

First we review the broad field of high power microwave emission, calculate the cost of Beacons, then consider radiating strategies for Beacons.

## 2. High Power Sources and Array Antennas

Since the early SETI era of the 1960s, microwave emission powers have increased orders of magnitude and new technologies have altered our ways of emitting very powerful signals. The highest *peak* power systems on Earth (peak powers over 10 GW) trade peak power for average power in order to get to a much stronger signal at distance at the lowest cost. There are also technical reasons why high power systems are pulsed, because 1) plasmas form inside sources and limit pulse duration 2) heating and cooling add cost and set limits on parameters such as duty factor.). Most of these high power devices operate in bursts of short pulses and for fundamental reasons are not extremely narrow band, having bandwidths of 0.01-1% (see Appendix B). Economical Beacons are likely to be pulsed: "The most rational ET signal would be a series of pulses that would be evidence of intelligent design." (Drake, 1990). This would be similar to the strategy of the lighthouse, pulsing and swinging the beam to get noticed.

The high power microwave (HPM) field–a mix of sources in the regime 0.1-100 GHz that either push conventional microwave device physics in new directions, or employ altogether new interaction mechanisms–has driven a huge leap in microwave power levels. HPM generation taps the enormous power and energy reservoirs of modern intense relativistic electron beam technology. Therefore, it runs counter to the trend in conventional microwave electronics toward miniaturization with solid-state devices, which are intrinsically limited in their peak power capability.

One definition of high power microwaves is
- devices that exceed 100 MW in peak power and
- span the centimeter- and millimeter-wave range of frequencies between 1 and 300 GHz.

This definition is arbitrary, and does not cleanly divide HPM and conventional microwave devices, which have, in the case of klystrons, exceeded 100 MW, HPM devices have reached output power in waveguide as high as 15 GW (Benford, 2007).

HPM is the result of the confluence of several historical trends. In the 1960s, electrical technology blossomed with the introduction of *pulsed power*, leading to the production of charged particle beams with currents in excess of 10 kA at voltages of 1 MV and more. The development of intense relativistic electron beams used the knowledge of wave-particle interaction gained in the study of plasma physics to be put to use in the generation of microwaves. The first HPM sources were descendants of conventional microwave sources such as the magnetron, the backward wave oscillator, and the traveling wave tube. In these devices, increased power comes from using higher operating currents and stronger beam-field couplings within the interaction region or on relativistic effects, most prominently the gyrotron. After 1970 came the introduction of new devices, such as the relativistic klystron, which depend fundamentally on the very high beam currents and strong space-charge forces that accompany high voltages.

A measure of the success of these efforts is the product of the peak microwave power and the square of the frequency, $Pf^2$. One meaning is that power density in the far field of a microwave beam transmitted from a fixed-size antenna is proportional to $Pf^2$ for a fixed antenna aperture (because antenna gain $\sim f^2$), so it is a good parameter to rank sources for Beacon power beaming over a distance. Figure 1 shows the general history of the development of microwave sources in terms of this factor. Conventional microwave devices ("tubes") made three orders of magnitude progress between 1940 and about 1970, but thereafter only minor improvements were made. HPM devices began at $Pf^2 \sim 1$ GW GHz$^2$, and have progressed upward an additional three orders of magnitude in the ensuing 20 years. The device with the highest figure of merit produced to date is the Free Electron Laser (FEL), with a radiated output power of 2 GW at 140 GHz .

Figure 2 shows the peak power generated by a representative sample of high power sources as a function of the frequency. HPM devices now operate at peak power 0.1-10 GW and single pulse energy up to 1 kJ. The ultimate limits on HPM source peak power are not well known; proposals have been made for single sources with 100 GW, 10 kJ outputs. For comparison, typical radar outputs are 1 MW, 1J. Combining the outputs in phase has been shown for most types of HPM sources, and so in principle one could generate a peak power of 100 GW from several sources.

To date, HPM sources operate at much shorter pulse lengths and duty factor— the product of the pulse length and the pulse repetition rate— factors than conventional tubes. Duty factors for HPM sources are currently $<10^{-5}$, whereas for conventional devices it varies from 1 (continuous operation, CW) down to about $10^{-4}$. However, HPM technology exists for duty factors $\sim 10^{-2}$ if applications require it in the future. HPM systems typically operate in bursts of pulses because burst-mode operation reduces the overall prime power requirement for long repetitive bursts. There is buildup of heat in the system due to inefficiencies. The thermal inertia timescale is $\sim 10^3$ sec. Operating longer than this time requires additional cost for cooling sub-systems.

Arrays of antennas are the only means of producing the large radiating areas ($\sim$km$^2$) that Beacons require. They also have high reliability and degrade gracefully, as loss of a few antennas does not mean failure. Arrays are widely used in radio astronomy receiving and are being planned for the new Deep Space Network (Bagri *et. al*, 2007).

Powerful systems can also be made in principle from millions of low power solid-state sources, printed circuit boards with a single chip doing signal processing and microwave generation. Each of them feed several sub-wavelength-sized antenna elements, and control the phase of each antenna element (Scheffer, 2005). Densely packed printed circuit boards and integrated antennas are not yet a mature technology. The number of elements increases with the square of both frequency f and antenna diameter D. The cost is approximately proportional to the number of elements, and so the capital cost varies as the product of the power P and the number of elements: $\$ \sim PD^2f^2$. Hence, high power devices coupled through dishes compete better for systems with large apertures and higher frequencies. Solid state operates best at lower frequencies, <5 GHz.

## 3. High Power Microwave Beacons

We investigate how Beacons can be built using HPM technology (see Appendix A for comments on laser METI). To make such large systems, groups of sources will drive an array of antenna elements, which could be dishes; there are other candidate antenna types. The microwave sources can be either power amplifiers or phase-injection-locked oscillators. In the amplifier case, a master oscillator drives a large number of power amplifiers (MOPA) and low-power phase shifters before each amplifier control beam direction, i.e., pointing is done by a customized phase shifting signal input to each module. Mechanical pointing of the dishes can also be used. In the oscillator–only case, the module consists of a number of oscillators, oscillating together in phase, and high-power phase shifters on the outputs of the modules do pointing control. Such high power shifters are the most difficult components in the system. The full system is built of modules of phase-locked sources, such as magnetrons, gyrotrons, and Cerenkov generators.

*3.1 Minimizing Beacon Costs*

We seek general relations for cost optimization; cost must be considered since Beacons will always compete with other demands.

The total cost $C_T$ of such large HPM systems is driven by two elements- *capital cost* $C_C$, divided into the cost of building the microwave source $C_S$ and the cost of building the radiating aperture $C_A$, and the *operating cost* $C_O$, meaning the operational labor cost and the cost of the electricity to drive the system:

$$\begin{aligned} C_T &= C_C + C_O \\ C_C &= C_A + C_S \end{aligned} \quad (1)$$

One can argue that operating cost of a system is dominated by labor cost, which is in turn proportional to the size and power as well, so that

$$C_C \propto C_O$$
$$C \propto C_C \quad (2)$$

Further, a rule-of-thumb for the cost of operating large facilities is about 10% of the capital cost to build it. Therefore optimizing the capital cost of building the Beacon also roughly optimizes the total cost $C_T$.

*Linear cost scaling*

To optimize, meaning minimize, the cost, the simplest approach is to assume power-law scaling dependence on the peak power and antenna area. This is a well-established method in industry (Benford *et al.*, 2007). Non-linear dependence is analyzed below, but to illustrate the essentials, first we analyze linear scaling with coefficients describing the dependence of cost on area a($/m$^2$), which includes cost of the antenna, its supports and sub-systems for pointing and tracking and phase control, and microwave power p ($/W) which includes the source, power supply, cooling equipment and prime power cost.

$$C_A = aA$$
$$C_S = pP \quad (3)$$
$$C_C = aA + pP$$

(We neglect any fixed costs, which would vanish when we differentiate to find the cost optimum. Mass production can decrease the cost of antenna elements and power modules, and is factored into the coefficients a and p.)

The power density S at range R is determined by W, the effective isotropic radiated power (EIRP), the product of radiated peak power P and aperture gain G,

$$W = PG$$
$$S = \frac{W}{4\pi R^2} \quad (4)$$

and gain is given by area and wavelength:

$$G = \frac{4\pi\varepsilon A}{\lambda^2} = \frac{4\pi\varepsilon A}{c^2} f^2 = kAf^2$$
$$W = kPAf^2 \qquad (5)$$
$$k = \frac{4\pi\varepsilon}{c^2}$$

where $\varepsilon$ is aperture efficiency (this includes factors such as phase, polarization and array fill efficiency) and we have collected constants into the factor k. (For $\varepsilon=50\%$, $k= 7\times10^{-17}$ $s^2/m^2$ and $kf^2=70$ for $f=1$ GHz.) We carry frequency as a constant with respect to optimization; cost of varying frequency is treated below. *EIRP determines the maximum range R of detectability of our Beacon* (see section 4). To find the optimum for a fixed power density at a fixed range meaning, from Eq. 4, fixed W, we substitute W into the cost equation,

$$C_C = \frac{aW}{kf^2 P} + pP \qquad (6)$$

then differentiate with respect to P and set it equal to zero, giving the optimum power and area:

$$\frac{\partial C_C}{\partial P} = -\frac{aW}{kf^2 P^2} + p = 0$$
$$P^{opt} = \sqrt{\frac{aW}{pkf^2}} \qquad (7)$$
$$A^{opt} = \sqrt{\frac{pW}{akf^2}}$$
$$A^{opt} = \frac{p}{a} P^{opt}$$

Optimal antenna diameter D is:

$$D^{opt} = \left[\frac{16}{\pi^2 f^2} \frac{pW}{ak}\right]^{1/4} \qquad (8)$$

The optimum (minimum) cost is

$$C_C^{opt} = \sqrt{\frac{aWp}{kf^2}} + \sqrt{\frac{aWp}{kf^2}} = 2\sqrt{\frac{aWp}{kf^2}}$$

$$C_C^{opt} = 2aA^{opt} = 2pP^{opt} \qquad (9)$$

$$\frac{C^{opt}{}_A}{C^{opt}{}_S} = 1$$

*Minimum capital cost is achieved when the cost is equally divided between antenna gain and radiated power.* This rule-of-thumb is used by microwave system designers for rough estimates of systems. It was first mentioned in 1968 (Brown, 1968), stated in the Project Cyclops report (Oliver and Billingham, 1996, pg. 61), and derived analytically (O'Loughlin, 1999), This cost ratio was independently discovered from cost data on the Deep Space Network (Benford, 1995). For a recent example, Kare and Parkin have built a detailed cost model for a microwave beaming system for a beam-driven thermal rocket and compared it to a laser-driven rocket. They find that, at minimum, cost is equally divided between the two cost elements (Kare, 2006).

Beacon parameters are described in sections 3.3 and 4. An immediate conclusion is that more powerful Beacons are more efficient: For a fixed power density S at range, cost scales as $W^{1/2}$, so from eq. 4 *cost scales only linearly with range R, not as $R^2$*. Radiating into the galactic disk, the number of stars scanned $N\sim R^2$ for ranges to ~ 1000 ly, slightly slower beyond that (Oliver, 1996). That means that more powerful Beacons will have economies of scale, the number of stars radiated to increasing as the square of cost: $N\sim C^{opt\,2}$. the cost per star scales as $1/N^{1/2}$.

*Power-Law Cost Scaling*

If we generalize Eq. 3 to give cost elements for aperture and microwave power a power-law dependence, i.e., $A^\alpha$, $P^\beta$, the optimum ratio of power cost to aperture cost is $\beta/\alpha$:

$$C_C = aA^\alpha + pP^\beta \qquad (10)$$

The optimal aperture and power are:

$$P^{opt} = \left[\frac{\alpha}{\beta}\frac{a}{p}\left(\frac{W}{kf^2}\right)^\alpha\right]^{\frac{1}{\alpha+\beta}}$$

$$A^{opt} = \frac{W}{kf^2 P} = \left[\frac{W}{kf^2}\right]^{\frac{\beta}{\alpha+\beta}}\left[\frac{\beta p}{\alpha a}\right]^{\frac{1}{\alpha+\beta}}$$

(11)

and total cost is

$$C_C^{opt} = a\left[\frac{W}{kf^2}\right]^{\frac{\alpha\beta}{\alpha+\beta}}\left[\frac{\beta p}{\alpha a}\right]^{\frac{\alpha}{\alpha+\beta}} + p\left[\frac{\alpha}{\beta}\frac{a}{p}\left(\frac{W}{kf^2}\right)^\alpha\right]^{\frac{\beta}{\alpha+\beta}}$$

(12)

$$\frac{C^{opt}_A}{C^{opt}_S} = \frac{\beta}{\alpha}$$

(13)

*The cost ratio depends on only the exponents.* The range of the cost ratio is small, about a factor of two (see below).

To transmit over galactic distances, the Beacon antenna will be far too large to be a single aperture, such as a dish. It will be an array of phased radiators, driven by power amplifiers or phase-locked oscillators. Present experience with large arrays is sparse; not enough studies exist to allow a detailed scaling. One case is a group of N nearby dishes of area $A_0$, each driven by an amplifier directed from a central control giving frequency and phase information. Then the total area is $A=NA_0$. Various combinations of N and $A_0$ can be used; cost will depend on the tradeoff between N and $A_0$. Cost data is available for differing scales of arrays, but with differing values of Ns and $A_0$s. (For example, one general result is that larger dishes require more expensive mounts and pointing sub-systems[1].) In general, cost will scale up with $A^\alpha$ and there will be N of them to make up a large array. So, we can write the cost as $C_A \propto NA_0^\alpha$. However, builders of arrays for radio astronomy express the resulting data without N as $C_A \sim A^\alpha$. To keep this form, we can write cost as

$$C_A \propto A^\alpha = (NA_0)^\alpha = N^{\alpha-1} NA_0^\alpha \tag{14}$$

---

[1] Typical scalings are aperture material cost~A, pointing cost~$A^{1/2}$, movement cost~$A^{3/2}$, stiffness cost~$A^2$.

The cost scaling of $NA_0^\alpha$ is therefore an overestimate by a factor of $N^{\alpha-1}$, but this becomes negligible if $\alpha \sim 1$, as is likely for advanced civilizations.

The exponent ranges from $\alpha=1$ to 1.375 (Bagri *et al.*, 2007). The Project Cyclops report concluded $\alpha=1$ (Oliver, 1996,) while *SETI 2020* gave $\alpha=1.35$ (Ekers et. al., 2002). It is difficult to argue that cost will vary less than linearly with area, so $\alpha=1$ is probably a minimum. For space-based antennas, with no requirement to support or move the antenna mass against gravity, $\alpha$ will be less than on a planet or moon. That argues that advanced space-faring civilizations further up the learning curve will approach $\alpha =1$.

For power costs, the rule-of-thumb in industry is $\beta=1$, and O'Loughlin gives $\beta=0.75-1$ (O'Loughlin, 1999). Therefore, in the range of Earth experience, the range of the cost ratio is small, about a factor of two:

$$\begin{aligned} 0.75 &< \beta < 1 \\ 1 &< \alpha < 1.375 \\ 0.55 &< \frac{C^{opt}_A}{C^{opt}_S} < 1 \end{aligned} \tag{15}$$

The coefficient a, microwave antenna cost per unit area, can be estimated from astronomical arrays, such as ALMA (National Research Council, 2005). Such systems are driven by tolerances for higher frequencies, ~100 GHz, and are ~4k\$/m$^2$. Proposed systems such as the Square Kilometer Array operate at lower frequencies near the minimum of attenuation, 1-10 GHz. SKA is projected to cost about 1 k\$/m$^2$ (Weinreb, 2001). Commercial dishes for satellite TV cost about \$400/m$^2$ and show the cost-lowering effects of huge economies of scale. SETI 2020 estimates ~\$350/m$^2$ for a large system (Ekers *et. al.* 2002). We have chosen 1 k\$/m$^2$ for our examples, but one must be mindful that the optimum area and power in Eq. 9 and 12 depend approximately on only the square root of the cost coefficients a and p, so are not very sensitive to changes in the technology.

Delivered microwave power currently costs ~$1/watt, including everything from the wall plug to the antenna connection. A current point of comparison, the International Thermonuclear Experimental Reactor (ITER) electron cyclotron heating system, heats the Tokamak with 27 gyrotrons of 1 MW continuous power (24 at 170 GHz, 3 at 120 GHz) at a projected cost of $82.5 million, $3/watt (Temkin, 200). This includes $26.3 million for power supplies, $14.5 million for gyrotron microwave sources and controls, and $41.7 million for waveguides transmission lines. We have chosen 3 $/W for our quasi-continuous Beacon examples.

Here we make an important distinction between peak and average power costs. The cost elements are different in pulsed peak power microwave systems, where cost is dominated by the requirement to operate at very high voltage and currents. The physics for the cost reduction is that the electrical breakdown threshold is much higher for short pulses, so much more energy can be stored in small volumes. For peak power systems, costs are in the range of 0.1-0.01$/W. Then a GW power system costs 10-100 M$. High average power systems have costs driven by continuous power-handling equipment and cooling for losses. They cost ~3$/W, so a GW unit costs ~3B$. Below we show the impact of average vs. peak power on Beacon cost (section 3.3) and performance (section 4).

*3.2 Cost Optimized Beacon Examples*

Cost optimization allows estimates of the area and power of Beacons. For an HPM system of W=EIRP=$10^{17}$ W, assuming $\alpha$=1, $\beta$=1, a=1k$/m$^2$, p=3$/W, at 1 GHz. We assume the aperture will be made of an array of elements with aperture efficiency $\varepsilon$=0.5. The cost becomes

$$C_C(B\$) = \frac{1.49}{P(GW)} + 3P(GW) \qquad (16)$$

It's expensive, costing billions of dollars. Figure 3a shows the sharp cost minimum. The cost of the Beacon falls rapidly as power increases until power cost equals aperture cost at total cost minimum of 4.14 B$. Then it increases monotonically. Optimum power is 0.69 GW, optimum antenna area is 2.07 km$^2$, and diameter is 1.62 km if the aperture is circular.

The difference between peak and average power costs has an impact on Beacon design. First, significantly lower costs for pulsed sources drives cost down, as shown in Fig. 3b, which compares the Beacon of Fig. 3a (which used 3$/W) with short-pulse Beacons with 0.3$/W and 0.03$/W. This is the known range of costs on Earth and shows an order of magnitude fall in cost, consistent

with Eq. 7 and 9.  Optimum power increases by ten to 7 GW, area falls by ten to 0.2 km². Below we show that the antenna area determines the angular beamwidth, so these smaller antennas have broader beams. That influences search strategy, as discussed in Section 4.4 and 5.

The scaling of aperture is very important for cost, as Figure 4 shows: more expensive antenna area ($\alpha$=1.375) drives cost upward. More power is radiated to make up the EIRP. The ratio of costs is

$$\frac{C^{opt}_A}{C^{opt}_S} = \frac{1}{1.375} \tag{17}$$

in agreement with Eq. 12. The main point of Figs 4 and 5 is that when antenna cost increases because $\alpha$=1.375, the total cost minimum increases to 20.7 B$, a more than six-fold rise. This drives the builder to increase power to 3.85 GW; optimum antenna area is 0.37 km², diameter is 0.69 km. There is a great incentive to keep $\alpha$ close to 1.

The EIRP for Beacons that operate across galactic scale is in the range of $10^{17}$-$10^{19}$ W or higher. As EIRP increases, optimal area, power and diameter must increase, as Fig. 5 shows for the constants of Fig. 3. As EIRP increases $10^{19}$ W, power rises to 7 GW, area to 20 km², and diameter to 5 km.

The exponents $\alpha$ and $\beta$ have a big impact on cost, but the cost dependence on $\alpha$ and $\beta$ is complex. Costs vs. radiating area is given in Fig. 6 for the practical ranges of $\alpha$ and $\beta$. Generally, lowering an index lowers cost. Figure 7 gives Beacon cost vs. power for a range of $\alpha$. Optimal cost increases with $\alpha$ and almost exactly linear with power. Figure 8 shows the rapid increase of cost with $\alpha$ seen in Fig. 7. Clearly, Beacon builders will move toward $\alpha$ =1.

We find the optimum total capital costs for a galactic-scale Beacon to be in the range of 10 $B. The Apollo project cost about 300 B$ in current dollars, and large science today, such as the Large Hadron Collider, International Linear Collider and ITER fusion reactor, are of order 10 B$. This suggests that galactic-scale Beacons are plausible luxuries.

*3.3 Frequency Choice*

Equations 9 and 12 show cost declining as frequency increases, due to the increased gain. The antenna cost also increases with frequency. But the increase is slow, and the coefficient a~$f^{1/3}$ (DeBoer, 2006). Microwave sources typically don't depend on frequency (power does, P~$f^2$, so more sources will be used to generate a given power at higher frequency). Therefore, the cost scaling with frequency is

$$C_c \propto \frac{1}{f^{5/6}} \qquad (18)$$

This is little different for the general case of Eq. 12, $C_C$~$f^{0.9}$. Figure 9 shows frequency dependence of cost for two of the cases of Fig. 6. For the linear case, cost drops a factor of 6.8 from 1 GHz to 10 GHz, so Beacons would more likely been seen at higher microwave frequencies.

The galactic background noise spectrum is flat between 1 GHz to 10 GHz. This is also the lowest attenuation region of earth's atmosphere (Ekers, et. al, 2002). The scaling of C with $1/f^{5/6}$ implies that in this flat region, the most favored spectral region is near 10 GHz, since this minimizes the cost of the Beacon (for any given range and hence EIRP), while imposing no noise cost on the receiver. (Note that this is quite different from some SETI thought, which privileges the "water hole" region between 1 and 2 GHz. Indeed, the metaphorical resonance between the spectral lines of H I and OH with "meeting at the water hole" may be a classic case of anthropic reasoning.) The secondary reasons given as early as Project Cyclops--that the low end of this band demands less stringent frequency stability-- vanishes if the Beacon is broadband, as we argue in Appendix B.

Since that era, detection of over 100 spectral lines in the interstellar medium, many of them organic, undermines the classic argument. Further, synchrotron radiation in the 1 GHz region increases going inward toward the galactic center, where the highest density of older stars peaks. A further benefit of higher frequencies for both Beacon and receiver: interstellar scintillation fades quickly with frequency, and can be ignored around and above 10 GHz (Cordes, 1991). As Beacon builders we will prefer that the listener not be confused by scintillation.

Our conclusion is that cost, noise and scintillation argue for radiating above the "water hole', especially if space-based. In the atmosphere, the optimum will be below 10 GHz where atmospheric attenuation minimizes.

*3.4 Beamwidth and Transmit Strategy*

The angular width θ of the optimized Beacon beam is set by the antenna area and frequency:

$$G = \frac{4\pi}{\theta^2}$$

$$\theta_o^2 = \frac{c^2}{A^{opt}f^2} = \frac{c^2}{f^2}\left[\frac{kf^2}{W}\right]^{\frac{\alpha+\beta-1}{\alpha+\beta}}\left[\frac{\alpha a}{\beta p}\right]^{\frac{1}{\alpha+\beta}} \quad (19)$$

where we have removed the aperture efficiency ($\varepsilon = 1$). For the linear case, $\alpha=\beta=1$,

$$\theta_o^2 = \frac{c^2}{f}\sqrt{\frac{ak}{pW}}$$

$$\theta_o = c\left[\frac{ak}{f^2 pW}\right]^{1/4} \quad (20)$$

The beamwidth from such a large cost-optimized aperture is $\theta_o \sim 10^{-4}$ radians. For example, the case of Fig. 3 gives $\theta_o = 2.5 \times 10^{-4}$ radians = 50 arc sec. Figure 10 shows the beamwidths for the Beacons of Fig. 6 as a function of EIRP. Note that beamwidth $\theta_o$ is higher for higher areal cost scaling $\alpha>1$, but depends weakly on EIRP, varying as $W^{-1/4}$ for the linear case and about the same, $W^{-0.28}$, for $\alpha, \beta \neq 1$.

For short-pulse, higher peak power Beacons, the beamwidth is substantially larger, $\theta_o \sim 10^{-3}$ radians. For example, the case of Fig. 3 gives $\theta_o = 2.5 \times 10^{-4}$ radians = 50 arc sec. The lower-cost case in Fig. 3b gives a factor of ten higher beamwidth, $\theta_o = 2.5 \times 10^{-3}$ radians = 500 arc sec = 8.33 arc min.

The primary implication of the small optimum emitting angle is that the transmission strategy for an Earth-based Beacon will be a rapid scan of the galactic plane, to cover the angular space. Such pulses will be infrequent events for the receiver. The *dwell time* when the Beacon beam falls on the receiver of a listener $\tau_d$ is related to the *revisit time* or cycle time $\tau_r$ that the Beacon with optimal emitting angle $\theta_o$ takes to broadcast across a segment of the galactic plane $A_G$. F is the fraction of the sky $A_G$ covers, as seen from the Beacon:

$$\frac{\tau_d}{\tau_r} = \frac{A(\Theta_o)}{A_G} = \frac{\pi\theta^2}{4A_G} = \frac{\pi\theta^2}{4(4\pi F)} = \frac{\theta^2}{16F} \tag{21}$$

For example, the galactic plane as seen from an Earth-based Beacon is typically ~10% of the sky, the case of Fig. 3 gives $\tau_d/\tau_r = 3.9 \times 10^{-8}$. If a Beacon builder wishes to cover the galactic plane in an Earth year, $\tau_r$ is $3.1 \times 10^7$ sec, $\tau_d$ is 1.2 sec. A receiver gets a short burst of pulsed microwaves, and does not see it again until a year later. Given the many possible local transient transmissions near a receiver (automobile spark plugs and other short-range machine timescales), a persistent signal for few seconds could be intuitively the best choice. A Beacon would linger a moment or two in our skies, and be back within something like a year. No search we know could have been likely to see such an event. Given the shortness of pulses of such a strategy, perhaps cost-optimized Beacons will be built to cover smaller, promising portions of the sky, and so revisit more often.

As beamwidth depends on the square root of $[a/pW]^{1/2}$, then 1) receive time decreases for expensive power, 2) receive time increases for cheap power, 3) receive time will be shorter for higher EIRP, meaning that the higher powers at a given range will be seen less frequently.

We can describe strategies to broadcast over long distances in the galaxy in a nomograph (Fig. 11), which is made from the beamwidth and times relations of Eq. 20-21. To use it, the Beacon builder, knowing his cost coefficients a and p, and choosing W (EIRP) from what range he wants (see next section), calculates $\theta_o$. Then he chooses a sky fraction F to illuminate, and locates that point on the $\theta_o$,F plain. This fixes $\tau_d/\tau_r$ from eq. 21, diagonal lines of constant duty cycle ratio, when the observer sees his Beacon. For example, for the case of Fig. 3, $\theta_o = 2.5 \times 10^{-4}$ rad and choosing F=0.1, $\tau_d/\tau_r = 3.9 \times 10^{-8}$, a line parallel to those at $10^{-8}$ and $10^{-6}$ in Fig. 11. Next, the builder chooses either a dwell or a repeat time and the other is fixed. If he chooses the revisit time to be a day, then from Fig. 11 he broadcasts to a single area in the sky for 3 ms.

The point is that choosing $\theta_o$ and F and moving along the dwell/revisit line is equivalent to choosing a search strategy. The cost-optimal Beacons lie in the upper region of the nomograph, and have a strategy of 'painting the sky' at long ranges, illuminating numbers of stars. In the lower right are the Beacons that many SETI searches have sought for, continuous (or near-continuous) Beacons that target one or a few stars, so can be seen by surveys that observe for short times.

## 4. Galactic Beacon Examples

To quantify some classes of possible Earth-based Beacons as we have discussed here, based on modern Earth sources, we work some examples. It's important to make as few assumptions as possible, but be specific as possible .

A Beacon of power $P_t$, frequency f, bandwidth $B_t$, antenna area $A_t$, produces a power density S (W/m$^2$) at range R (Eqs. 4, 5)

$$S = \frac{P_t k A_t f^2}{4\pi R^2} \qquad (22)$$

The receiver of antenna area $A_r$ and receiver bandwidth $B_r$ collects a power

$$P_r = W \frac{A_r}{4\pi R^2} \frac{B_r}{B_t} \qquad (23)$$

for $B_r < B_t$. For $B_r \geq B_t$, $B_r/B_t = 1$.

This signal power competes with noise power $P_n$ in the receiver, given by the Nyquist relation, determined by the receiver system temperature $T_{sys}$:

$$P_n = k_B T_{sys} B_r$$
$$\frac{P_r}{P_n} = \frac{WA_r}{4\pi R^2 k_B T_{sys} B_t} \qquad (24)$$

where $k_B$ is the Boltzmann constant. Integrating over a time t in the bandwidth channel $B_r$, the noise is reduced, giving a signal-to-noise ratio (Kraus, 1986)

$$\frac{S}{N} = \frac{P_r}{P_n} \sqrt{B_r t} = \frac{WA_r}{4\pi R^2 k_B T_{sys} B_t} \sqrt{B_r t} \qquad (25)$$

The transmitter bandwidth $B_t$ is set by microwave source physics and is of order 1 MHz. (This quite different from the assumption of early SETI searches, where $B_t$ was hoped to be ~1 Hz, to allow easier detection. But microwave device physics makes high power devices have higher bandwidths, as shown in Appendix B.)

Now we consider the effect on Beacon design using continuous, slowly-pulsed near- continuous vs. short-pulse technology. Parameters for the three examples are shown in the Table. The fixed parameters are: Beacon and receiver antennas we assume to be of equal area, $\alpha=1$, $\beta=1$, f=1 GHz, $\varepsilon=0.5$, $T_{sys}=7K$, bandwidth of transmitter $B_t$=1 MHz.

Example 1: Quasi-continuous Galactic Range Beacon To show the implications of choosing short- or long-pulse technology and its impact on listeners, we base this Beacon on gyrotron technology, which is commercially available today, operating in 1-second pulses in bursts at 0.5 Hz, so the duty factor~1 (DF=product of pulse length and repetition rate), which we call 'long-pulse'. We choose the case of Figure 3, but for higher EIRP=$10^{19}$ W and fixed parameters as given above. From fig. 5, total power is 6.9 GW, perhaps generated by about three thousand present state-of-art gyrotrons, from an antenna array of diameter 5.1 km. Gain is ~$10^9$, comparable to the largest Earth antennas. We choose to illuminate 1% of the sky, ~400 deg$^2$, which encompasses `10% of the galaxy seen from Earth (Sullivan, & Mighell, 1984). The revisit/dwell time ratio is 2.7x$10^{-8}$. For one year revisit, $\tau_d$ is 0.8 sec, for one month, 0.07 sec.

At what range might such a Beacon be seen? Such sources have bandwidths of ~1 MHz (Appendix B), so if the listener is looking for this class of Beacon with such bandwidth with integration time 1 sec, $(B_r t)^{1/2}$~$10^3$. Assuming a space-based receiving array in outer regions of an alien star system, where system temperature is determined by zodiacal light, $T_{sys}$ = 7 K, we need then to assume the area of the radio telescope. Assuming the same order of investment as ourselves, we take the same diameter as the Beacon, 5.1 km. (Large arrays in space are more likely to be built with $\alpha$~1 because of lack of loading on the structure.) Power density received in 1 MHz bandwidth is 0.078 J. To achieve S/N=5, the range for seeing this Beacon is 6,080 light-years. Thus, an EIRP of $10^{19}$ W is a truly galactic-scale Beacon. As R~$P_t^{1/2}$ (eq. 25), to reach Galactic Center from Earth, 28,000 light-years, requires EIRP~$10^{20}$ W, as in the third example.

Note that to quasi-continuously beam into the whole sky (isotropic) would cost 10 times as much.

Example 2: Short-pulse Medium-Range Beacon This is an 'entry-level' Beacon. The approach is to build a short-pulse technology Beacon, with cheaper power. Fig. 3b shows that short-pulse sources on Earth are cheaper by a factor of 10-100 vs. long-pulse sources, and here we assume 100, p=0.03$/W. From Eqs. 7 and 9, we get higher power, smaller diameter and lower Beacon cost (1.3 B$), larger beam angular size and longer dwell time for fixed revisit time.

A major difference comes in the factor $(B_r t)^{1/2}$ in Eq. 25. The duty factor of pulsed sources reduces the advantage of integration for the alien observer because integration occurs only while the pulse is on, not between pulses. HPM sources currently have duty factors <$10^{-5}$. However, technology exists for duty factors ~$10^{-2}$ if applications require it. In this example, we assume the Beacon will be built with such advanced sources. To compare to the first example, the difference to the observer will be the pulse length; Example 1 is a series of 1-sec pulses, Example 2 a rapid series of 1 μsec pulses in a burst.

To be specific, base the Beacon on a master oscillator driving an array of power amplifiers driving dish antennas. In operation, it fires a burst of short pulses, then by phase adjustment repositions the beam to the next location in the sky. If the observer optimizes his electronics for a wide variety of bandwidth and pulse duration, by providing data bins of all sizes (1 Hz, 1 kHz, 1 MHz, etc.), for frequencies in the microwave, then pulsed sources can be observed with $(B_r t)^{1/2} \sim 1$. Making the same assumptions for equal transmitter and receiver antenna area and $T_{sys}$, and keeping S/N=5, range falls to 1,900 ly as cost falls.

<u>Example 3: Short-pulse Galactic Range Beacon</u> We can extend the range by increasing EIRP from the second example, scaling up short-pulse technology with 10 times more modules for 10 times more powerful Beacon, $W=10^{20}$ W, increasing power to 218 GW, and diameter to 2.9 km. This approach has 13.1 B$ cost, lower than the long-pulse Beacon. The range is longer, 19,227 ly, most of the distance to the galactic center.

The result of short-pulse technology is to give cheaper, longer-range Beacons. Note that these costs will fall if these Beacons operate at 10 GHz, so that the short-pulse, medium-range example drops to 0.19 B$ (Eq. 18, Fig. 9).

## 5. Implications of Cost Optimization for Beacon Broadcast Strategy

At distances >1000 light years, with many stars in the field of view, Doppler adjustment to offset relative motions, becomes pointless. Further, distortion of signals from >1000 light years arises from interstellar absorption and scintillation. Such "twinkling" of the signal comes from both the dispersion of differing frequencies, and delays in arrival time for pulses moving along slightly different pathways, due to refraction. Temporal broadening probably would limit bandwidth to >1 MHz, as we know from the broadening of pulsar signals. To get more information into a pulse, a Beacon could use a technique known as Wavelength Division Multiplexing, which yields a broad signal with subsections, each hundreds of kilohertz broad. The downside of looking for broadband Beacons of $\Delta f > 10$ MHz is the greater chance of confusing these with natural emissions, such as the 1420 MHz hydrogen emission. Rapid pulses, though, will stand out against the steady background. Since they are periodic, they will not look like scintillation; if they are at higher frequencies around 10 GHz, they will have little scintillation.

*5.1 The Galactic Habitable Zone*

There is a growing opinion within the astrobiological community that we live among the outer regions of a Galactic Habitable Zone (Gonzales *et.al.*, 2001; Trimble, 1997). Lineweaver *et.al.* argue that early, intense star formation toward the inner Galaxy provided the heavy elements necessary for life, but the supernova frequency remained dangerously high there for several billion years. Later, stars orbiting between the crowded inner bulge and the barren outer Galaxy were born into a habitable zone, starting about 8 Gy ago. The habitable zone expanded with time as metalicity (driven by supernovas) spread outward in the Galaxy and the supernovae rate decreased. They argue that ~ 75% of the stars that harbor complex life in the Galaxy are older than the Sun with average age ~ 1 Gy older than the Sun, which implies that most advanced societies should lie much farther inward toward the galactic center, at distances > 1000 light years. Broadcasting to relatively near stars misses most of the possible civilizations.

*5.2 Galactic Center Broadcast Strategy*

Our Beacons, as we envision them, should broadcast into the plane of the spiral disk. From Earth, 90% of the galaxy's stars lie within 9% of the sky's area, in the plane and hub of the galaxy, so we should follow a limited sky broadcast strategy. Special attention should be paid to areas along the Galactic Disk. The observers will need to be patient and wait for recurring events that arrive in periodic bursts. To make the revisit more frequent, several Beacons can be constructed or we can broadcast into a small sector with frequent returns to attract attention, when switch to another small sector.

Whatever races might dwell further in from us toward the center, they must know the basic symmetry of the spiral. This suggests the natural corridor to communicate with them is along the spiral's radius. (A radius is better than aiming along a spiral arm, since the arm curves away from any straight-line view of view. On the other hand, along our nearby Orion arm the stars are roughly the same age as ours.) This avenue maximizes the number of stars within a Beacons view, especially if we broadcast at the galactic hub. Thus, a Beacon should at least broadcast radially in both directions. Radiating into the full disk takes far more time and power; such Beacons will rarely visit any sector of the plane. Of course such distances imply rather larger Beacons.

We can summarize the implications of these cost-conscious results as five transmission strategies:

1. Scan the entire plane of the galaxy often. Broadcast toward and away from the galactic center often, perhaps daily.

2. Since the highest nearby density of stars lies along the nearby Orion galactic arm, broadcast toward that direction occasionally.

3. Broadcast toward the locations of the transient, powerful bursts seen in past SETI surveys in a systematic way, over times ~ year (Benford, G. *et al.*, 2009).

## 6. Conclusions

The evolution of high power microwave sources in the half-century since the founding ideas of SETI has given us new, high power technologies. We applied these to consider how galactic-scale Beacons could be built, driven by cost constraints, on Earth or in space in our solar system. The ratio of antenna area and microwave transmitting power costs on Earth ranges only over a factor of two. Further, cost declines with frequency, due to the increased antenna gain. Cost-efficient beacons will be pulsed, narrowly directed, and broadband ($\Delta f/f$ ~0.1%) in the 1-10 GHz region, with a cost preference for the higher frequencies. Cost, spectral lines near 1 GHz and interstellar scintillation favor radiating far from the "water hole." We've given the scaling and examples of such Beacons.

Transmission strategy for such Earth-based Beacons will be a rapid scan of the galactic plane, to cover the angular space. Such pulses will be infrequent events for the receiver, appearing for only seconds, recurring over periods of a month or year.

Our discussion has many implications for the reverse problem, SETI itself. If other civilizations are thrifty and minimize costs, what would their Beacons look like and how should we search for them? This is the subject of our second paper (Benford, G., *et al*., 2009).

## 7. Acknowledgements
We acknowledge helpful comments from Jill Tarter, Seth Shostak, Carlo Kopp, Paul Shuch, Steven Baxter, Geoff Landis, Paul Davies and Richard Gott.

## Author Disclosure Statement

No competing financial interests exist.

## Appendix A: Laser METI

The cost optimization arguments in this paper are not specific to any frequency domain. Laser Beacons are a plausible alternative to the microwave (Howard *et al.*, 2004). There are different issues, however.

We have concentrated on long range Beacons, which in microwaves can target areas of the sky. Lasers within ~1000 ly can target single solar systems. In fact, laser frequencies are so high that diffraction-limited beams can target the Habitable Zone of individual stars out to ~1000 ly: for a 7-m aperture and 1 μm wavelength, the spot size at 10000 ly is 10 AU. This is the most attractive laser Beacon broadcast strategy, because beyond about 1000 ly, maximum energy in a laser pulse drops by 40%, and falls further exponentially with distance; this means 2 magnitudes reduction at 3000 ly. While lasers may be preferred for targeting oxygen-atmosphere planets within a 1000 ly or so, they are not good choices in the Beacon range beyond, because beyond 1000 ly scattering and extinction makes targeting oxygen-bearing planets difficult. Moving from the far infrared (10 μm) helps, but lasers' advantage in picking out life-bearing stars vanishes in just the range that microwave beacons become best.

Microwaves are now a mature technology, making plausible the cost parameters. But laser METI is too young to allow a meaningful cost calculation. Project Ozma occurred in the year lasers were invented, 1960, and laser technology is still rapidly changing. There are no clear-cut choices for preferred wavelengths. Large (> 1-m) high-quality mirrors for laser systems and high power laser costs are difficult to estimate. There are only a few devices in existence, compared to microwaves, where many systems exist. Typical costs are roughly a~1M$/m$^2$, p~10$/W (Kare & Parkin, 2006). These cost coefficients imply that laser Beacons will be small in aperture, high in power.

**Appendix B: Why Does Bandwidth Increase With Power?**

The typical way to generate very high powers in to build up an oscillation in a cavity and release it in a pulse. Energy in a cavity is the integral of the electric field energy over the volume, which can be extracted in Q cycles, Q being the cavity quality factor. The number of modes N possible in a cavity is proportional to the volume in units of wavelength $\lambda^3$:

$$P = \frac{\omega}{Q} \int_V \frac{1}{2}\varepsilon E_{rf}^2 dV,$$

$$N \propto \frac{V}{\lambda^3},$$

$$P \propto \frac{\lambda^2 E_{rf}^2 N}{Q}$$

(26)

(Note that power extracted scales with $\lambda^2$, as in Fig. 2.)

Higher power can be most easily produced by lowering Q. Q is also related to the bandwidth $\delta f$, $Q = f/\delta f$. In the latter case, $P \sim \delta f$, and thus higher power cavity devices have larger bandwidths than lower power cavities.

Less attractive methods are by increasing the electric field in the cavity, which is limited by breakdown strengths between surfaces, or by increasing the number of modes, which leads to 'mode competition', making cavity design more complex.

Zaitsev, A., 2006, "Messaging to Extraterrestrial Intelligence", http://arxiv.org/abs/physics/0610031

**Table**
**Cost-Optimized Galactic Beacons**
(For f=1 GHz, a=1k$/m$^2$, Beacon and receiver antennas of equal area,
$\alpha$=1, $\beta$=1, $\varepsilon$=0.5, $T_{sys}$=7K, bandwidth of transmitter $B_t$=1 MHz.)

| Beacon Parameter | Long-Pulse Galactic-Range Beacon | Short-Pulse Medium-Range Beacon | Short-Pulse Galactic-Range Beacon |
|---|---|---|---|
| **Range for S/N=5** | 6080 ly | 1080 ly | 10,800 ly |
| **EIRP=W** | $10^{19}$ W | $10^{18}$ W | $10^{20}$ W |
| **Power, $P^{opt}$** | 6.9 GW | 21.9 GW | 218 GW |
| **Antenna Diameter, $A^{opt}$** | 5.1 km | 0.91 km | 2.88 km |
| **Beamwidth, $\theta$** | $1.2\ 10^{-4}$ rad | $6.4\ 10^{-4}$ rad | $2.1\ 10^{-4}$ rad |
| **Pulse Length** | 1 s | 1 µs | 1 µs |
| **Repetition Rate** | 0.5 Hz | 1 kHz | 1 kHz |
| **Total capital Cost, $C_C$** | 41.4 B$ | 1.3 B$ | 13.1 B$ |
| **Power Cost Coefficient, p** | 3$/W | 0.03$/W | 0.03$/W |
| **Dwell Time, $\tau_d$** | 1.1 s | 35 s | 1.1 s |
| **Revisit Time, $\tau_r$** | 1 yr | 1 yr | 1 yr |

**Figure Legends**

Fig. 1 Growth of microwave device output in terms of quality factor $Pf^2$.

Fig. 2. Peak Powers of various types of HPM devices. Quality factor $Pf^2$ varies over orders of magnitude. Devices are Magnetically Insulated Line Oscillator (MILO), Backward wave Oscillator (BWO), Traveling Wave Tube (TWT), Free Electron Laser (FEL).

Fig. 3a. Antenna, microwave power and total costs of Beacon of EIRP=$10^{17}$ W, $\alpha$=1, $\beta$=1, a=1 k$/m$^2$, p=3 $/W, at f=1 GHz with aperture efficiency $\varepsilon$=0.5. Minimum total cost is at the point where the antenna cost and power cost are equal, as in Eq. 9

Figure 3b. Impact of pulsed sources on system cost of Beacons of EIRP=$10^{17}$ W. Case of long-pulse sources at 3$/W (same as Fig. 3 case) and short-pulse sources at 0.3$/W and 0.03$/W.

Fig. 4. Beacon of Fig. 3 with $\alpha$=1.375. More rapid cost scaling for antenna leads to smaller antenna, higher power and significantly higher total cost to achieve the same EIRP. Antenna and power cost are in ratio of $\beta/\alpha$, as in Eq. 13.

Fig. 5. Optimal area, diameter and power of Beacons as a function of effective isotropic radiated power, an extrapolation of case of Fig. 3.

Fig. 6. Beacon costs as a function of radiating area for various values of $\alpha$, $\beta$ and EIRP=$10^{17}$ W, a=1 k$/m$^2$, p=3 $/W, at f=1 GHz with aperture efficiency $\varepsilon$=0.5. The expected ranges of the antenna and power indices are shown, and hence the acceptable region is the v-shaped region enclosed by them. Cost falls with lower indices. Since the cost scale is logarithmic, the penalty for not constructing near the minimum is severe. Minimum cost is at an antenna diameter of hundreds of meters, and hence is large enough to suggest a phased array approach to keep structures to manageable scales.

Fig. 7. Beacon cost vs. power over the expected range of $\alpha$ for EIRP=$10^{17}$ W, $\beta$=1, a=1 k$/m$^2$, p=3 $/W, at f=1 GHz . Optimal cost increases with $\alpha$ and almost exactly linearly with power as $\alpha$ increases.

Fig. 8. Rapid increase of cost with $\alpha$. Compared with the previous figure this shows that the cost increases very rapidly with alpha, with a power law index of 7.2. This scaling will apply approximately to other (non-SETI) applications,

hence there will be pressure to develop technical approaches to building antennas that scale with $\alpha \approx 1$. For EIRP=$10^{17}$ W, a=1 k$/m$^2$, p=3 $/W, at f=1 GHz.

Fig. 8. Rapid increase of cost with $\alpha$. Compared with the previous figure this shows that the cost increases very rapidly with alpha, with a power law index of 7.2. This scaling will apply approximately to other (non-SETI) applications, hence there will be pressure to develop technical approaches to building antennas that scale with $\alpha \approx 1$. For EIRP=$10^{17}$ W, a=1 k$/m$^2$, p=3 $/W, at f=1 GHz.

Fig. 9. Cost vs. microwave frequency for two values of $\alpha$. The upper end of frequencies are favored by almost an order of magnitude, in contrast to the "waterhole" frequencies, which are between 1 and 2 GHz. Optimal cost (as desired by the Principle of Parsimony) drives the Beacon builder to higher frequencies.

Fig. 10. Optimal angular beamwidth $\theta_o$ of Beacon beam in radians (left) and arcseconds (right) for a large range of effective isotropic radiated power and for the Beacons of Fig. 6. Beamwidth depends weakly on EIRP, $\sim W^{-1/4}$, hence the probable beam areas fall into a narrow range. It should be noted that $\theta_o$ is larger for a higher areal cost index (large aperture cost implies smaller apertures, so larger beams), and for a lower power cost index (lower power cost implies smaller apertures).

Fig. 11. Nomograph for Beacon broadcast strategy. A Beacon builder choosing values of $\theta_o$ (from Fig. 10) and a sky fraction F to illuminate gives lines of constant duty cycle ratio (dwell time/revisit time) for the Beacon observer. Then right and top axes give ranges of these times for fixed $\tau_d/\tau_r$ ratio, and can not be correlated to the lower and left axes. The two relations are independent of each other, except in that they produce the same time ratios. Cost-optimal Beacons lie in the upper region, continuous Beacons targeting specific star targets are in lower region, can be observed with surveys observing for short times. The galactic-scale Beacon examples from the Table are in the upper right quadrant at F=0.01.

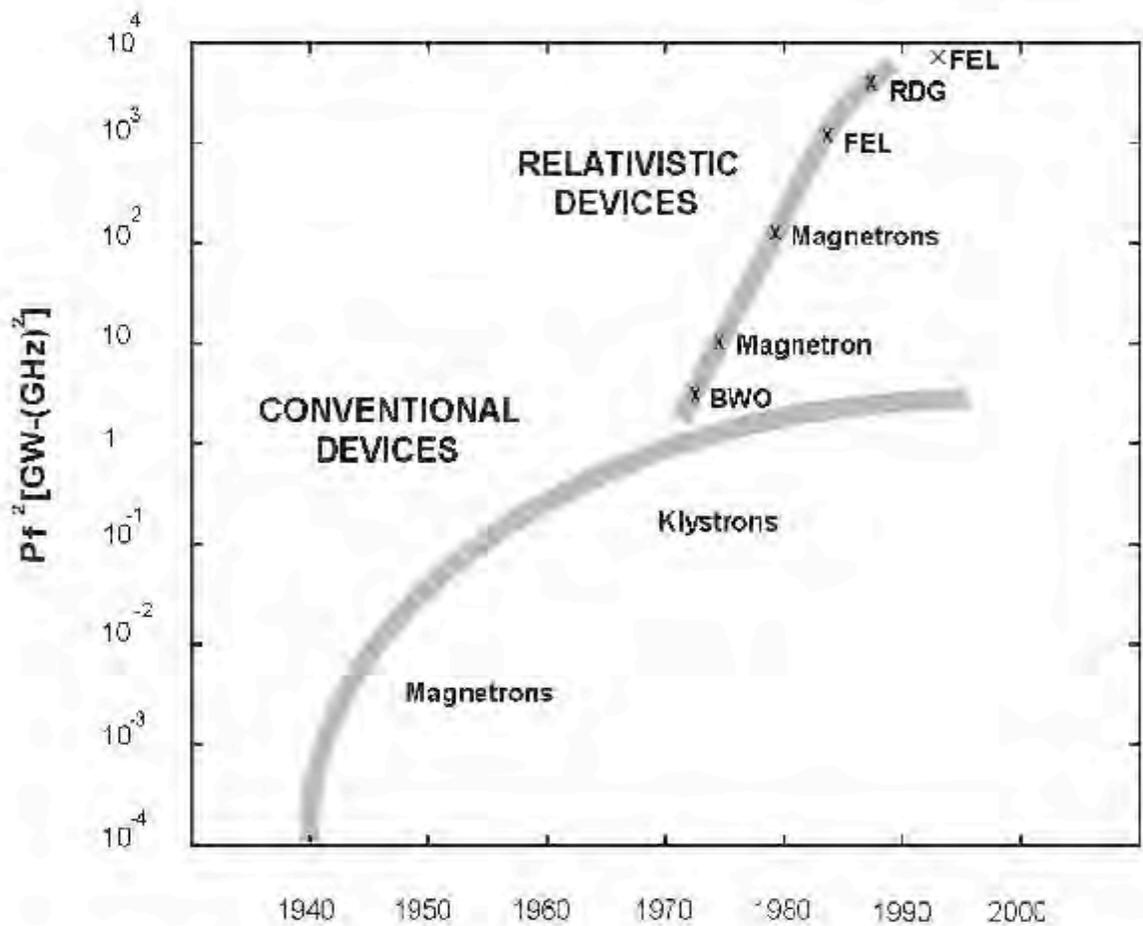

Fig. 1 Growth of microwave device output in terms of quality factor $Pf^2$.

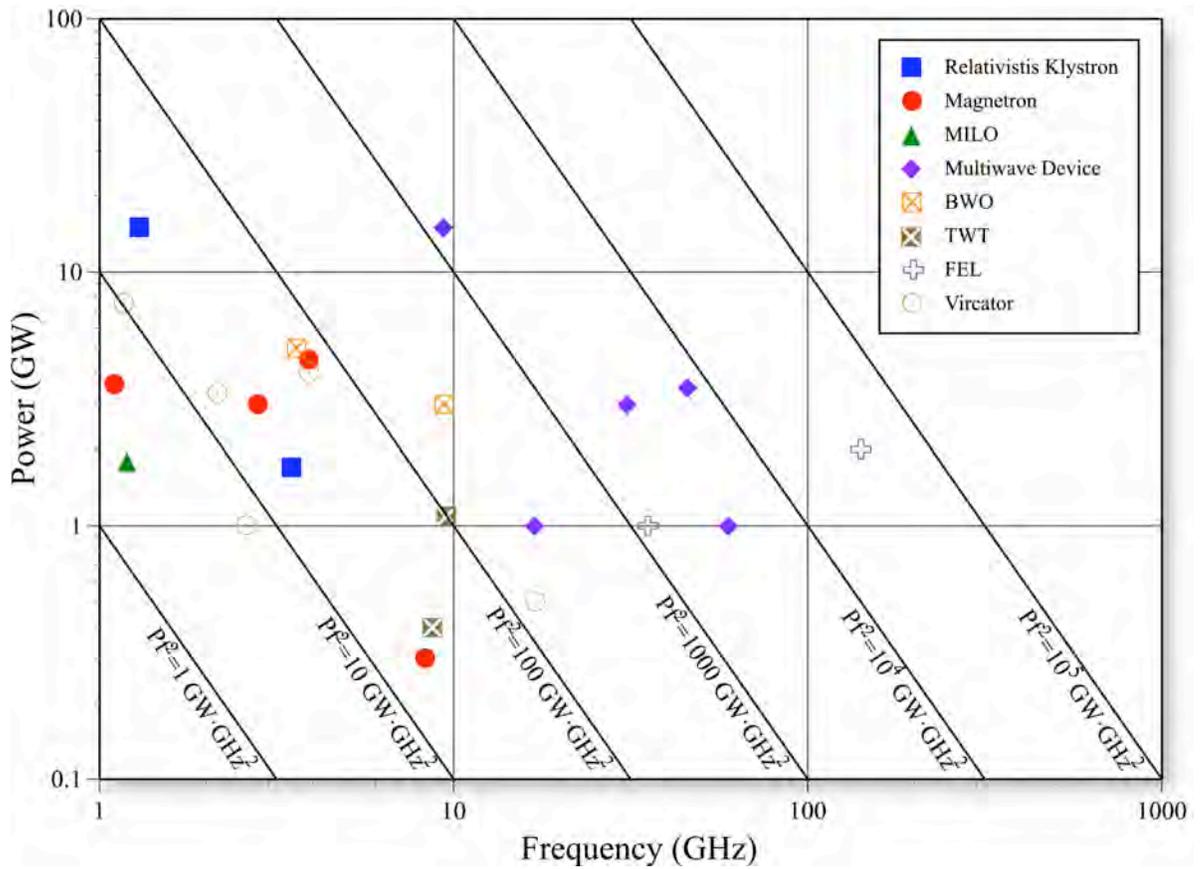

Fig. 2. Peak Powers of various types of HPM devices. Quality factor $Pf^2$ varies over orders of magnitude. Devices are Magnetically Insulated Line Oscillator (MILO), Backward wave Oscillator (BWO), Traveling Wave Tube (TWT), Free Electron Laser (FEL).

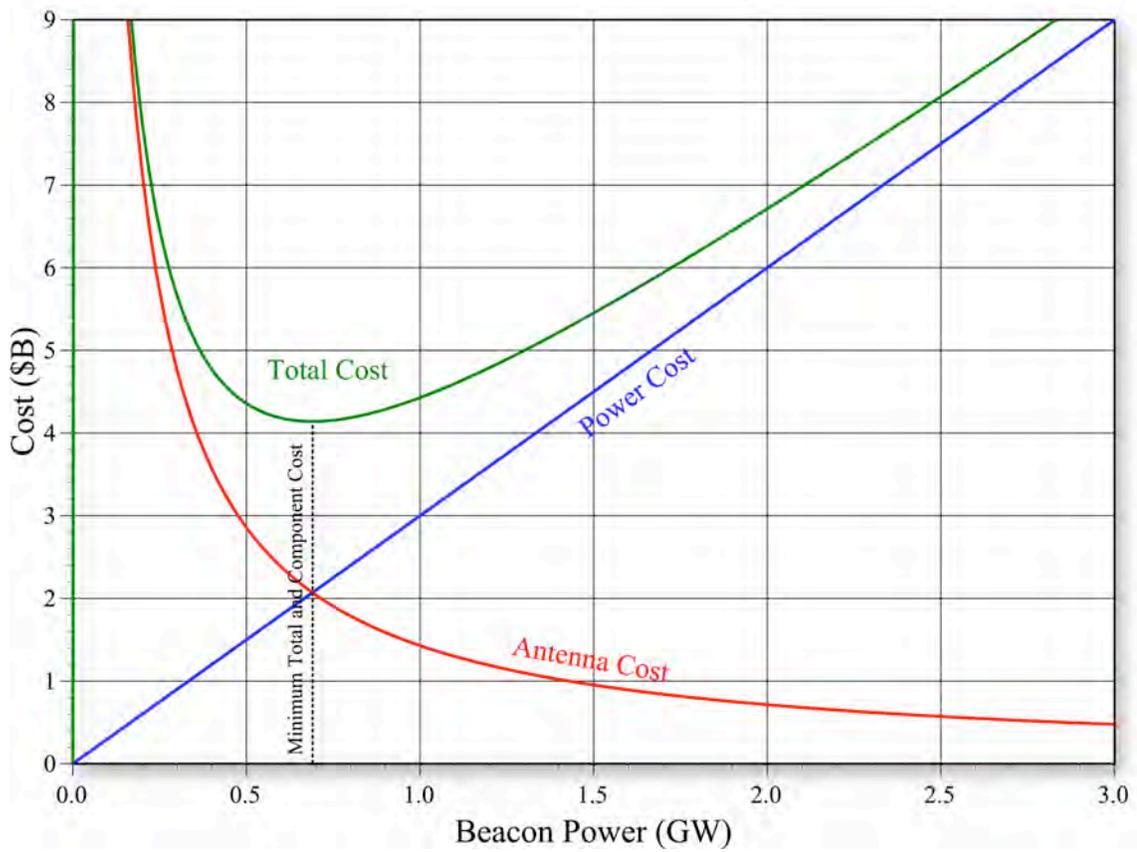

Fig. 3a. Antenna, microwave power and total costs of Beacon of EIRP=$10^{17}$ W, $\alpha=1$, $\beta=1$, $a=1$ k$/m$^2$, $p=3$ $/W, at $f=1$ GHz with aperture efficiency $\varepsilon=0.5$. Minimum total cost is at the point where the antenna cost and power cost are equal, as in Eq. 9.

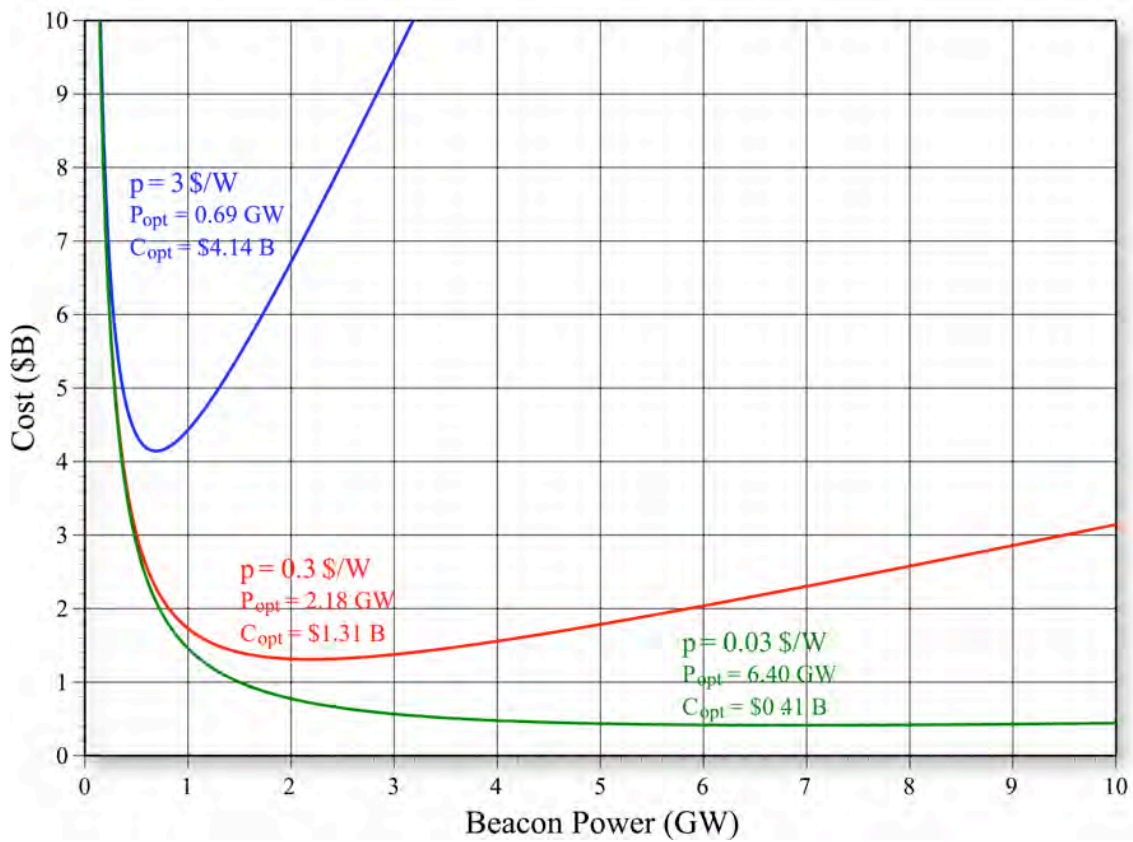

Figure 3b. Impact of pulsed sources on system cost of Beacons of EIRP=$10^{17}$ W. Case of long-pulse sources at 3$/W (same as Fig. 3 case) and short-pulse sources at 0.3$/W and 0.03$/W.

*

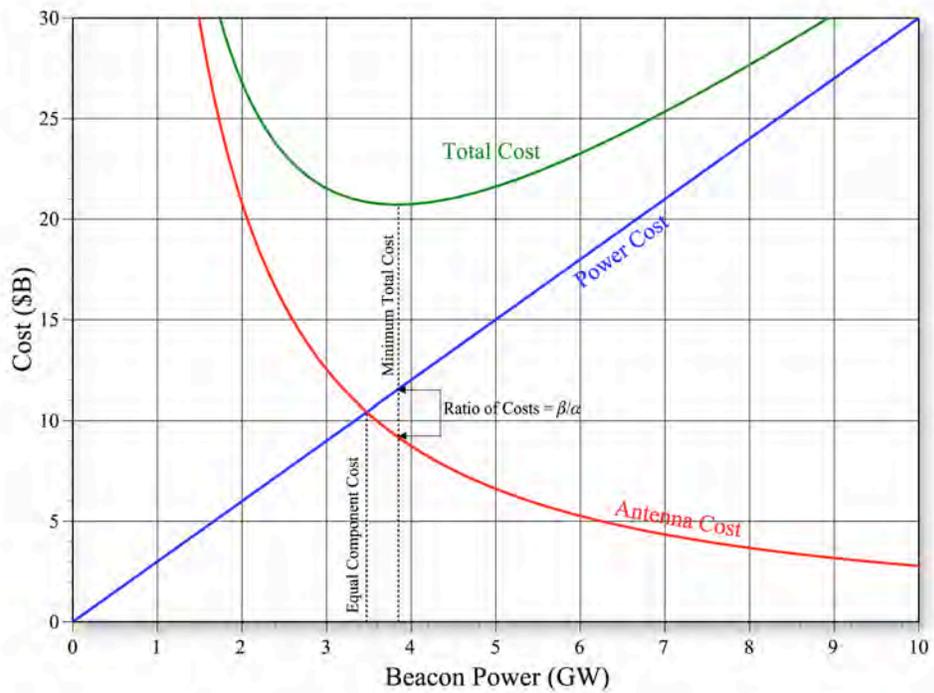

Fig. 4. Beacon of Fig. 3 with $\alpha=1.375$. More rapid cost scaling for antenna leads to smaller antenna, higher power and significantly higher total cost to achieve the same EIRP. Antenna and power cost are in ratio of $\beta/\alpha$, as in Eq. 13.

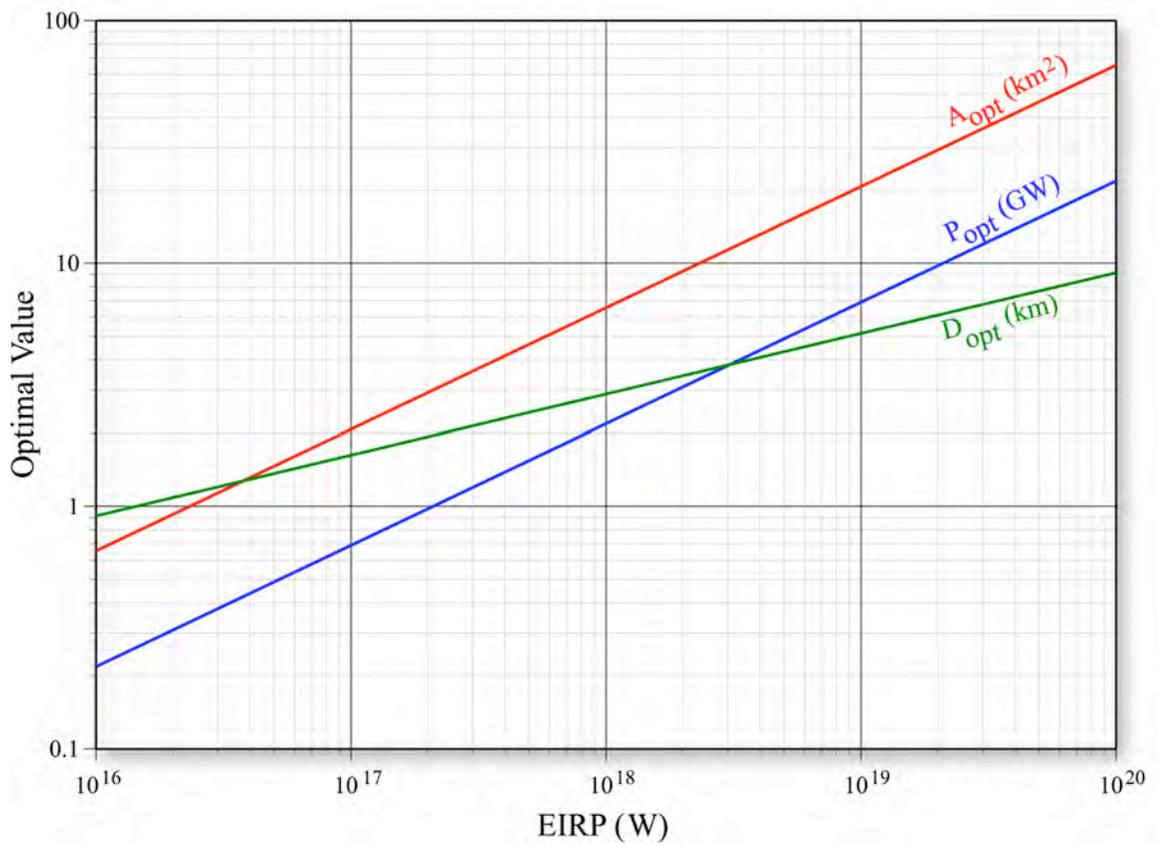

Fig. 5. Optimal area, diameter and power of Beacons as a function of effective isotropic radiated power, an extrapolation of case of Fig. 3.

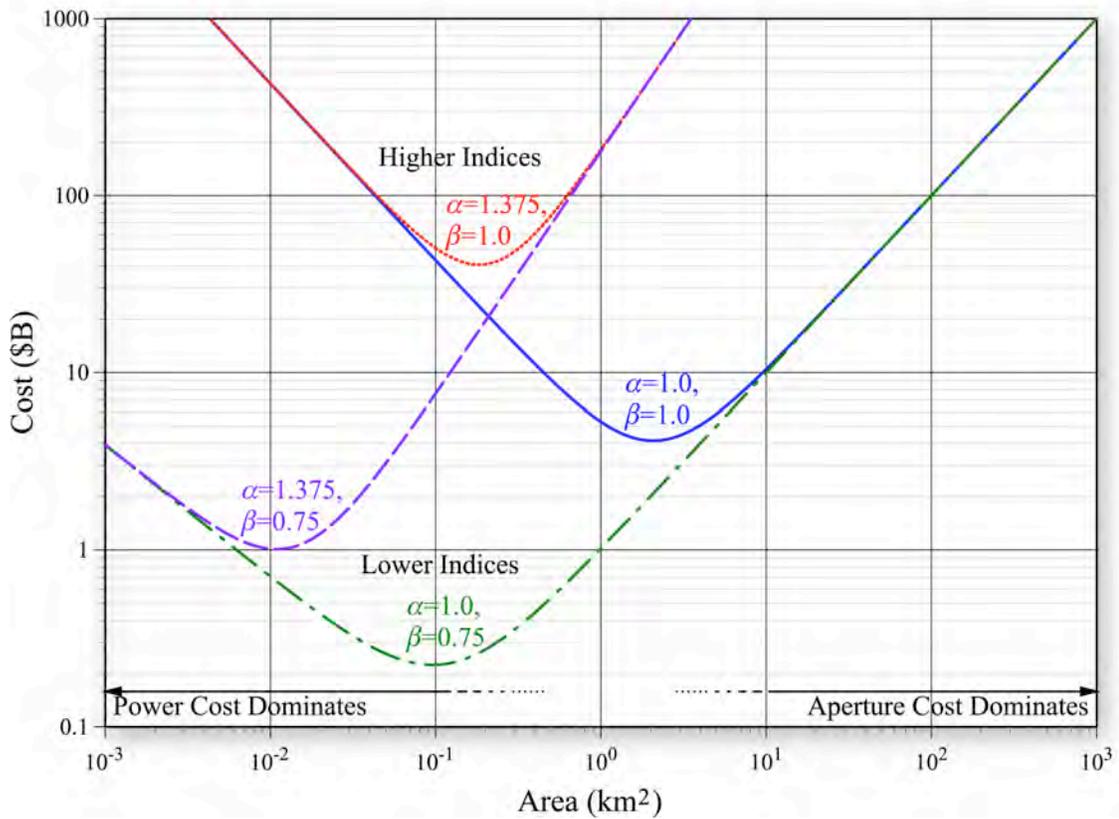

Fig. 6. Beacon costs as a function of radiating area for various values of α, β and EIRP=$10^{17}$ W, a=1 k$/m$^2$, p=3 $/W, at f=1 GHz with aperture efficiency ε=0.5. The expected ranges of the antenna and power indices are shown, and hence the acceptable region is the v-shaped region enclosed by them. Cost falls with lower indices. Since the cost scale is logarithmic, the penalty for not constructing near the minimum is severe. Minimum cost is at an antenna diameter of hundreds of meters, and hence is large enough to suggest a phased array approach to keep structures to manageable scales.

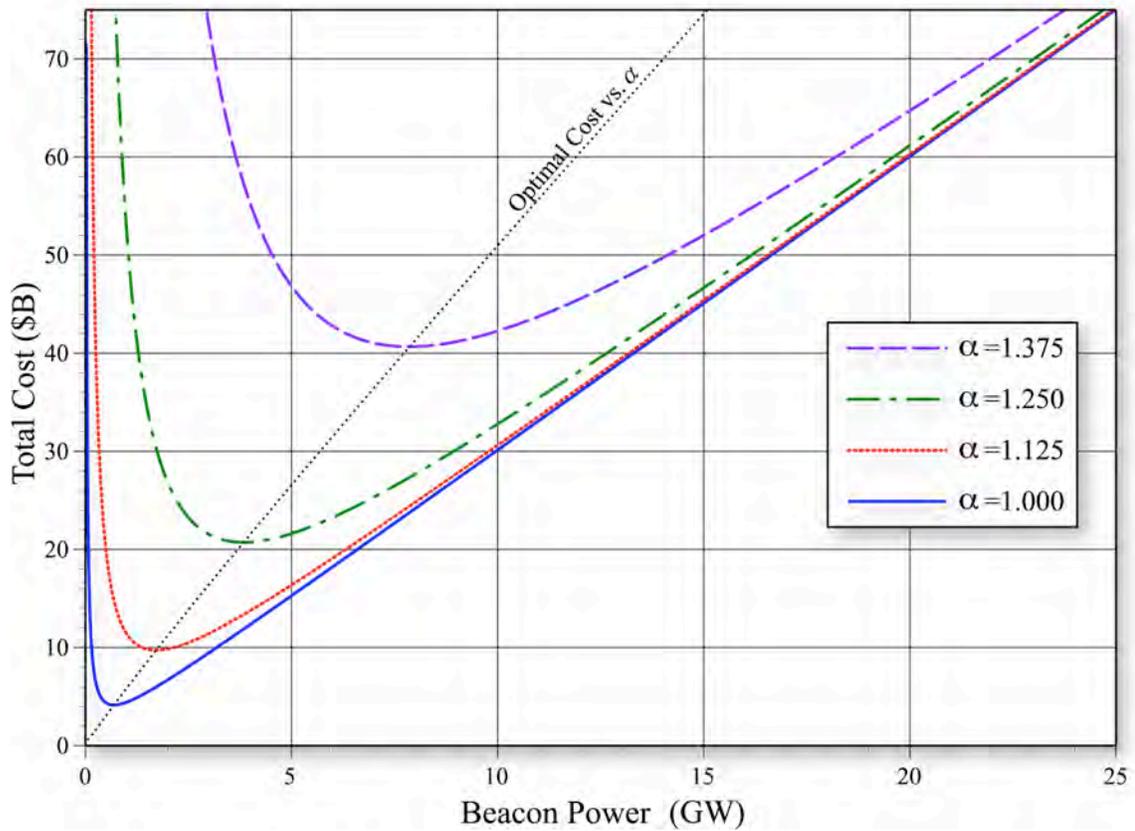

Fig. 7. Beacon cost vs. power over the expected range of $\alpha$ for EIRP=$10^{17}$ W, $\beta$=1, a=1 k\$/m$^2$, p=3 \$/W, at f=1 GHz. Optimal cost increases with $\alpha$ and almost exactly linearly with power as $\alpha$ increases.

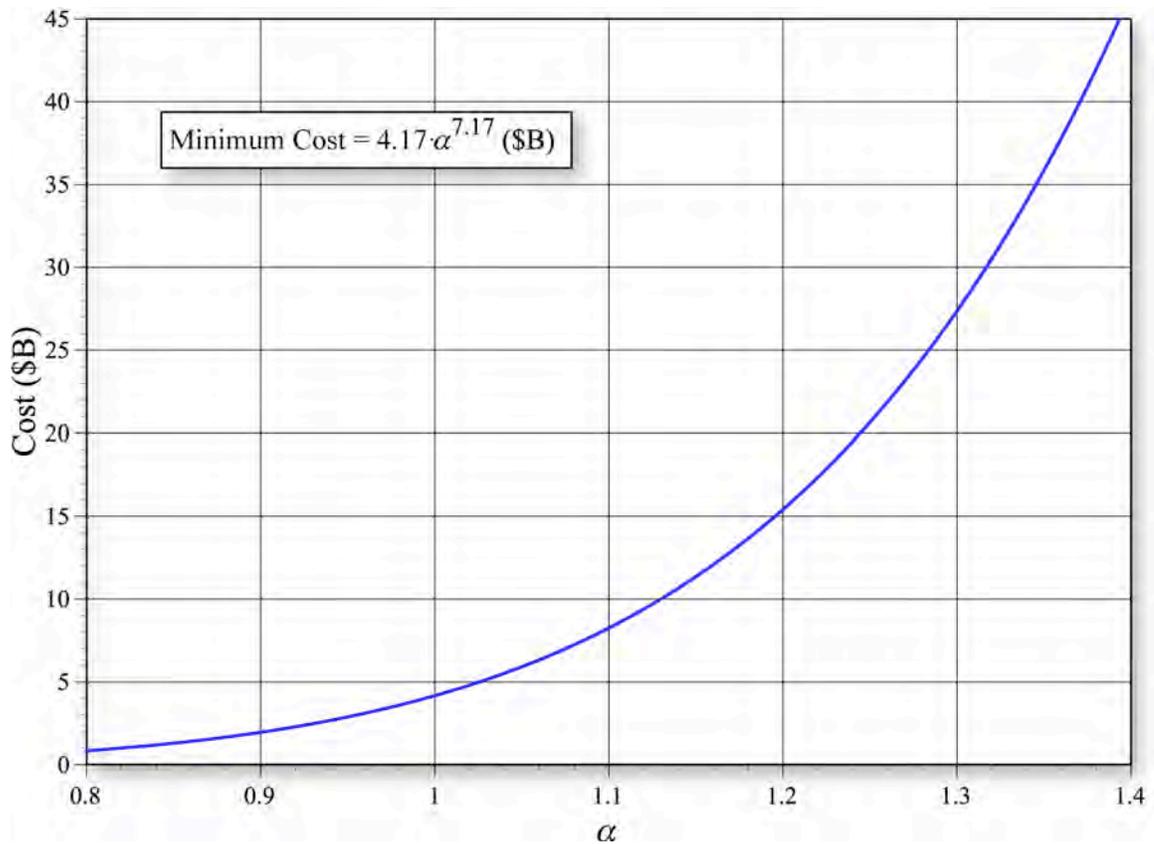

Fig. 8. Rapid increase of cost with $\alpha$. Compared with the previous figure this shows that the cost increases very rapidly with alpha, with a power law index of 7.2. This scaling will apply approximately to other (non-SETI) applications, hence there will be pressure to develop technical approaches to building antennas that scale with $\alpha \approx 1$. For EIRP=$10^{17}$ W, a=1 k$/m$^2$, p=3 $/W, at f=1 GHz.

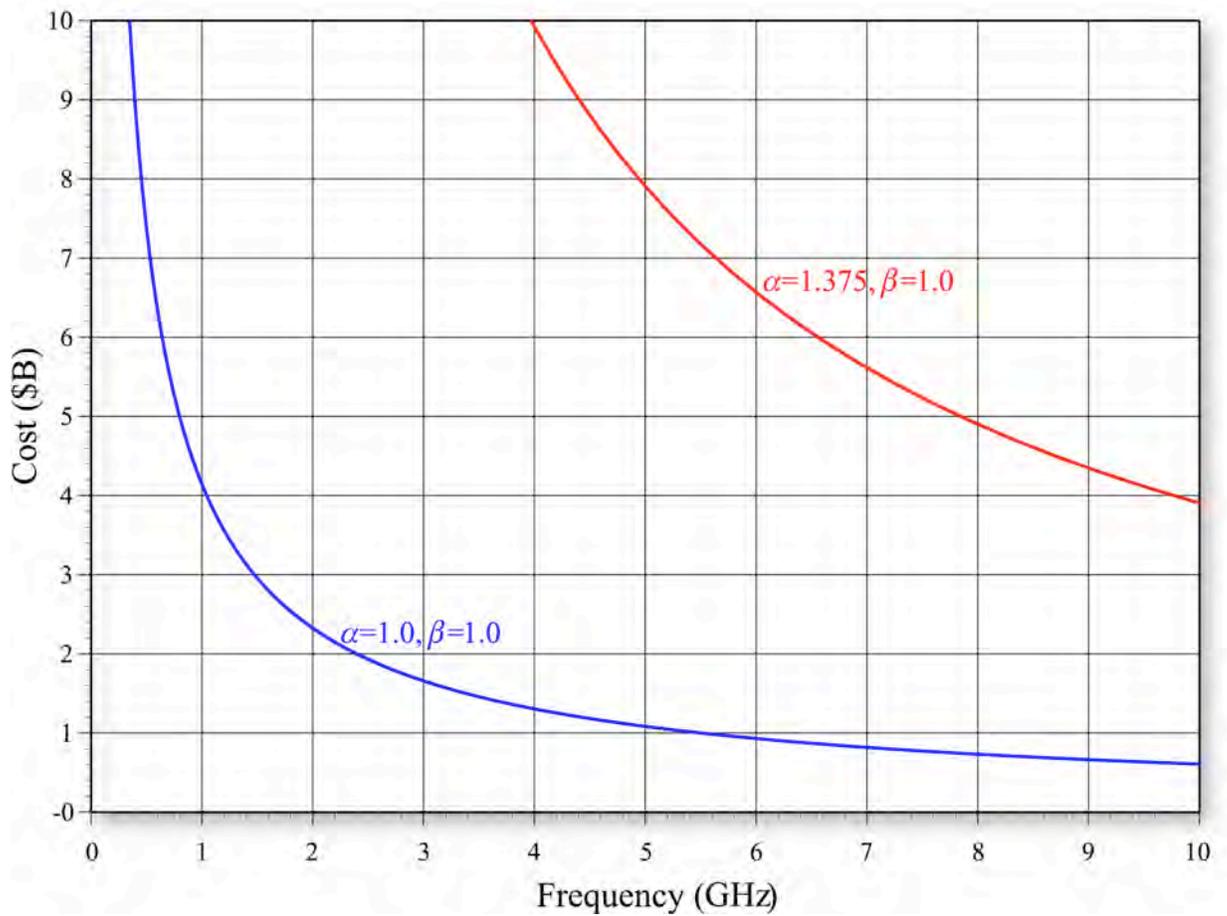

Fig. 9. Cost vs. microwave frequency for two values of α. The upper end of frequencies are favored by almost an order of magnitude, in contrast to the "waterhole" frequencies, which are between 1 and 2 GHz. Optimal cost (as desired by the Principle of Parsimony) drives the Beacon builder to higher frequencies.

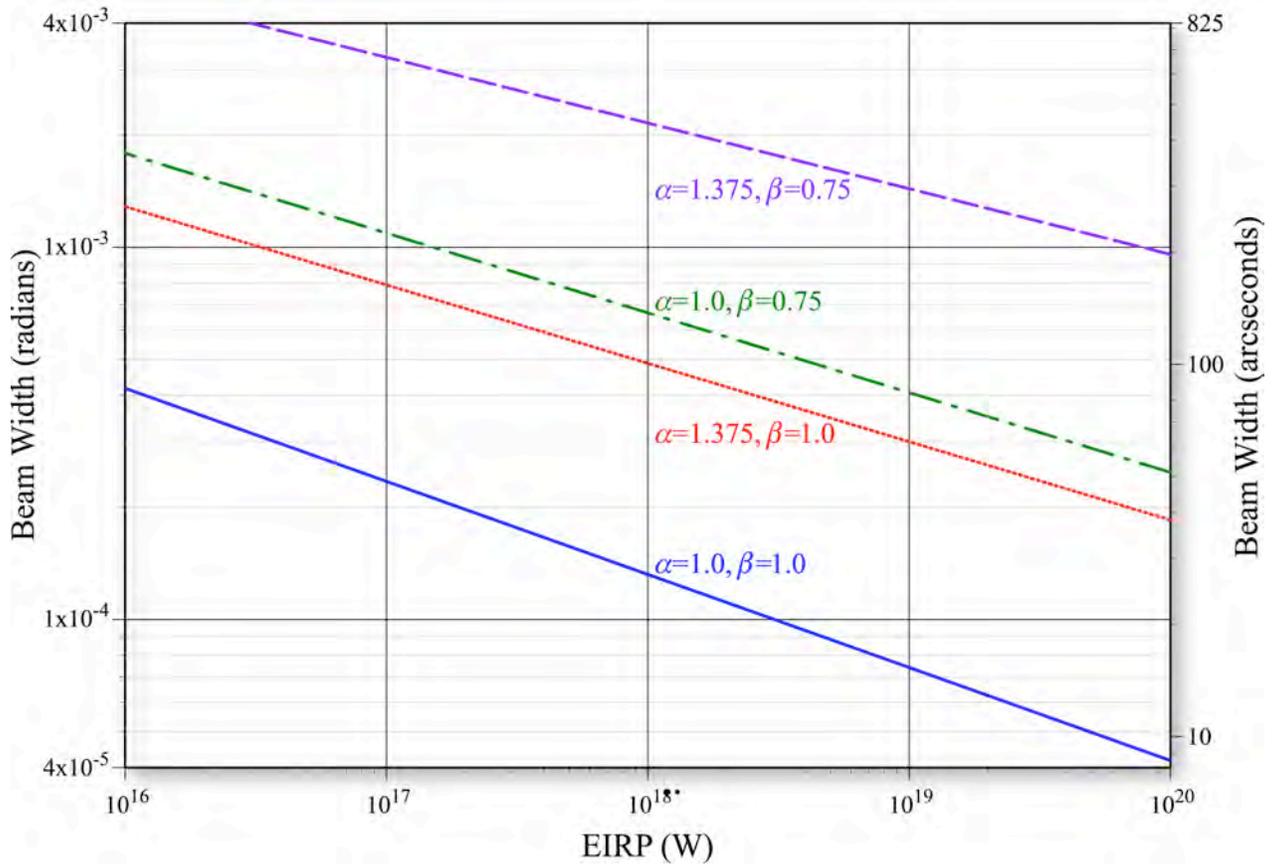

Fig. 10. Optimal angular beamwidth $\theta_o$ of Beacon beam in radians (left) and arcseconds (right) for a large range of effective isotropic radiated power and for the Beacons of Fig. 6. Beamwidth depends weakly on EIRP, $\sim W^{-1/4}$, hence the probable beam areas fall into a narrow range. It should be noted that $\theta_o$ is larger for a higher areal cost index (large aperture cost implies smaller apertures, so larger beams), and for a lower power cost index (lower power cost implies smaller apertures).

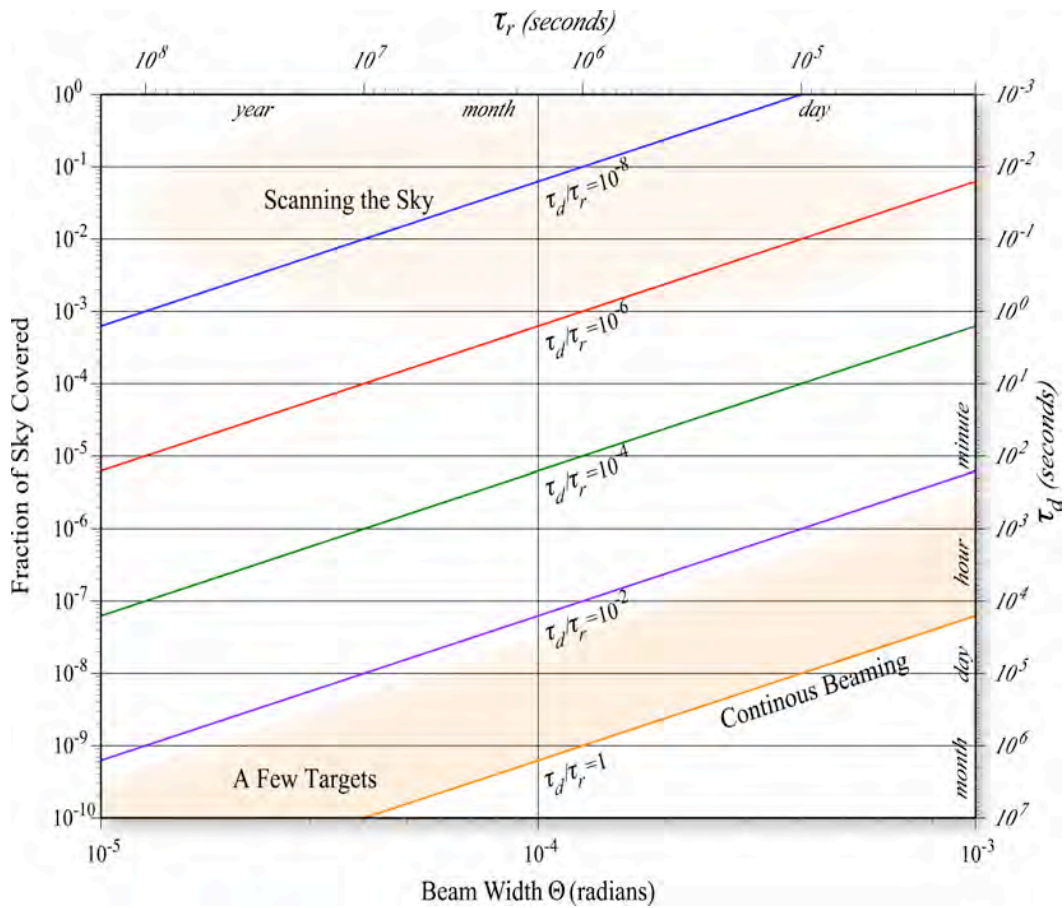

Fig. 11. Nomograph for Beacon broadcast strategy. A Beacon builder choosing values of $\theta_o$ (from Fig. 10) and a sky fraction F to illuminate gives lines of constant duty cycle ratio (dwell time/revisit time) for the Beacon observer. Then right and top axes give ranges of these times for fixed $\tau_d/\tau_r$ ratio, and can not be correlated to the lower and left axes. The two relations are independent of each other, except in that they produce the same time ratios. Cost-optimal Beacons lie in the upper region, continuous Beacons targeting specific star targets are in lower region, can be observed with surveys observing for short times.